%% file: main.tex
\documentclass[sigconf,nonacm]{acmart}

\input{tex/preamble}

\input{tex/copyright}

\begin{document}

\title{Adversarial Observations in Weather Forecasting}

\input{tex/authors}

\begin{abstract}
\input{tex/00_abstract}
\end{abstract}

\maketitle
\enlargethispage{-1cm}
\section{Introduction}
\label{sec:introduction}
\input{tex/01_introduction}

\newpage
\section{Weather Forecasting}
\label{sec:background}
\input{tex/02_background}

\section{Adversarial Observations}
\label{sec:method}
\input{tex/03_method}

\section{Evaluation}
\label{sec:eval}
\input{tex/04_evaluation}

\subsection{Concealing Extreme Predictions}
\label{sec:case_study}
\input{tex/05_case_study}

\section{Statistical Detection}
\label{sec:countermeasures}
\input{tex/06_countermeasure}

\section{Discussion}
\label{sec:discussion}
\input{tex/07_discussion}

\section{Related Work}
\label{sec:related-work}
\input{tex/08_related_work}

\section{Conclusion}
\label{sec:conclusion}
\input{tex/09_conclusion}

\bibliography{
    bib/acsac.bib,
    bib/asiaccs.bib,
    bib/ccs.bib,
    bib/raid.bib,
    bib/eurosp.bib,
    bib/icse.bib,
    bib/ndss.bib,
    bib/sp.bib,
    bib/uss.bib,
    bib/pldi.bib,
    bib/nips.bib,
    bib/additional.bib
}

\end{document}

%% file: tex/preamble.tex
\usepackage{microtype}

\usepackage{pgfplots}
\pgfplotsset{compat=1.18}
\usepgfplotslibrary{fillbetween}
\usetikzlibrary{intersections}
\usetikzlibrary{positioning}
\usepgfplotslibrary{statistics}
\usetikzlibrary{matrix}
\usepgfplotslibrary{groupplots}
\usetikzlibrary{plotmarks}

\usepackage{amsmath}
\usepackage{siunitx}
\usepackage{bbm}

\usepackage[scaled=0.825]{beramono}
\usepackage[T1]{fontenc}

\usepackage{xcolor}
\usepackage[]{hyperref}
\hypersetup{
  colorlinks,
  linkcolor={red!40!black},
  citecolor={red!40!black},
  urlcolor={red!40!black}
}
\usepackage[nameinlink,capitalise,noabbrev]{cleveref}
\crefname{lstlisting}{listing}{listings}
\Crefname{lstlisting}{Listing}{Listings}

\usepackage{listings}
\usepackage[linesnumbered,ruled,vlined]{algorithm2e}
\crefname{algocf}{alg.}{algs.}
\Crefname{algocf}{Algorithm}{Algorithms}

\usepackage[font=normal,labelfont=bf]{caption}
\usepackage{subcaption}

\usepackage{adjustbox}

\usepackage{tabularx, booktabs}
\usepackage{multirow}

\usepackage{ifthen}
\usepackage[inkscapearea=page]{svg}

\usepackage{array}

\lstdefinestyle{sexp}{
    language=Lisp,
    showstringspaces=false,
    basicstyle=\color{codeDarkColor}\ttfamily,
    keywordstyle=\color{codeColor}\ttfamily,
    stringstyle=\color{red}\ttfamily,
    identifierstyle=\color{codeColor},
    mathescape=true
}

\definecolor{poscolor}{RGB}{128, 180, 230}
\definecolor{negcolor}{RGB}{252, 90, 40}
\definecolor{neucolor}{RGB}{1, 74, 26}

\definecolor{color4}{RGB}{152, 190, 87}
\definecolor{color5}{RGB}{11, 49, 66}
\definecolor{gridcolor}{RGB}{85, 85, 85}

\usepackage{pifont} 
\usepackage{xspace}

\usepackage{enumitem}

\newcommand{\tikzcircle}[2][red,fill=red]{\tikz[baseline=-0.5ex]\draw[#1,radius=#2] (0,0) circle ;}
\newcommand{\tikztriangle}[2][red]{\tikz[baseline=-0.5ex]\node[#1] at (0,0) {\pgfuseplotmark{triangle*}};} 

\newcolumntype{R}[1]{>{\raggedleft\arraybackslash}p{#1}}

\definecolor{myred}{HTML}{bf616a}
\newcommand{\red}[1]{\textcolor{myred}{#1}}

\newcommand{\weatherstate}{\mathbf{X}}
\newcommand{\advloss}{\mathcal{A}}
\newcommand{\perturbation}{\boldsymbol{\delta}}
\newcommand{\inferenceFunction}{f}
\newcommand{\denoisingStep}{d}
\DeclareMathOperator*{\argmin}{arg\,min}

\makeatletter
    \def\balanceissued{unbalanced}
    \let\oldbibitem\bibitem
    \def\bibitem{%
        \ifnum\thepage=\PreviousTotalPages%
            \expandafter\ifx\expandafter\relax\balanceissued\relax\else%
                \balance%
                \gdef\balanceissued{\relax}\fi%
            \else\fi%
        \oldbibitem}
\makeatother

%% file: tex/copyright.tex
\ccsdesc[500]{Computing methodologies~Machine learning}
\ccsdesc[500]{Security and privacy~Software and application security}

\keywords{Adversarial Machine Learning, Weather Forecasting, Adversarial Robustness, Security of AI}

%% file: tex/authors.tex
\author{Erik Imgrund\vspace{0.1em}}
\affiliation{%
  \institution{BIFOLD \& TU Berlin}
  \country{Germany}
}
\author{Thorsten Eisenhofer\vspace{0.1em}}
\affiliation{%
  \institution{BIFOLD \& TU Berlin}
  \country{Germany}
}
\author{Konrad Rieck\vspace{0.1em}}
\affiliation{%
  \institution{BIFOLD \& TU Berlin}
  \country{Germany\vspace{1em}}
}
\renewcommand{\shortauthors}{Imgrund et al.}

%% file: tex/00_abstract.tex
AI-based systems, such as Google's GenCast, have recently redefined the state of the art in weather forecasting, offering more accurate and timely predictions of both everyday weather and extreme events. While these systems are on the verge of replacing traditional meteorological methods, they also introduce new vulnerabilities into the forecasting process.
In this paper, we investigate this threat and present a novel attack on autoregressive diffusion models, such as those used in GenCast, capable of manipulating weather forecasts and fabricating extreme events, including hurricanes, heat waves, and intense rainfall. The attack introduces subtle perturbations into weather observations that are statistically indistinguishable from natural noise and change less than $0.1\,\%$ of the measurements---comparable to tampering with data from a single meteorological satellite.
As modern forecasting integrates data from nearly a hundred satellites and many other sources operated by different countries, our findings highlight a critical security risk with the potential to cause large-scale disruptions and undermine public trust in weather prediction.

%% file: tex/01_introduction.tex
Weather forecasting plays a central role in our daily life, ranging from choosing appropriate clothing to managing critical operations in industry. Accurate forecasts, for instance, are essential for the operation of renewable energy systems, agricultural planning, aviation operations, and disaster risk mitigation.
In recent years, weather forecasting has seen significant advances, with AI-based approaches rapidly progressing and now beginning to surpass traditional numerical weather prediction~\citep{gencast,graphcast,aardvark}.

Currently, the leading system in this space is \emph{GenCast}~\cite{gencast}, an autoregressive diffusion model developed by Google. GenCast outperforms the best traditional medium-range forecasting system, ENS~\cite{ens}, in both day-to-day accuracy and the prediction of extreme weather events.
Due to these advances, major meteorological institutions, such as the US National Oceanic and Atmospheric Administration (NOAA) and the European Centre for Medium-Range Weather Forecasts (ECMWF), are preparing to incorporate AI-based approaches into their forecasting systems~\cite{aifs, ai-noaa}. With the frequency and intensity of extreme weather events increasing in recent years, this integration also represents a critical step toward more effective disaster risk mitigation on a global scale.

However, this shift also introduces a new security risk. Weather forecasting systems depend on observational data aggregated from a diverse array of organizations, each operating under different jurisdictions and guided by distinct institutional incentives~\cite{satellites}. Moreover, the underlying data sources are equally varied, encompassing land stations, weather balloons, aircraft, ships, and satellites~\cite{eyre2022assimilation}. This decentralized and fragmented data ecosystem creates a broad attack surface, offering adversaries multiple opportunities to tamper with observations.
The potential consequences of such manipulation are severe. Reliable weather warnings, for example, are indispensable for mitigating harm by enabling timely preparation and evacuation ahead of extreme events~\cite{benefitOfEarlyWarnings}.

In this paper, we explore the risk of manipulating AI-based weather forecasting systems. In particular, we introduce an attack for creating \emph{adversarial observations}, subtle changes to measurements that mislead the predictions of a weather model. While our approach is inspired by prior techniques for generating adversarial inputs~\citep{madryPGD,dp-attacker}, it addresses a key challenge specific to weather models based on autoregressive diffusion, such as GenCast. These models denoise and condition their input over multiple iterations, making standard gradient calculation technically infeasible and limiting the applicability of existing attacks. To overcome this challenge, we propose a novel approximation of the inference procedure that enables the computation of effective perturbations, capable of inducing false weather forecasts, such as fabricating non-existing extreme events or concealing real ones.

The core idea of our approach is to sample the inference process of a forecasting model at a tractable number of steps and iteratively estimate its gradient in reverse. Our approximation uses progressively smaller noise levels in each diffusion step, balancing the difficulty of the attack by including both small and large noise levels which stabilizes the optimization procedure. To ensure that all changes remain within acceptable bounds, we apply a projection operator tailored towards weather observations, which constrains each measurement variable individually based on its variance. 
As weather observations naturally exhibit variance, this projection ensures that the calculated perturbations remain indistinguishable from other sources of noise, such as measurement inaccuracies or inference errors.

To analyze the efficacy of this attack, we conduct an empirical evaluation across a broad range of geographic locations and time periods, using GenCast as the target model. Specifically, we construct adversarial observations to induce extreme events at specific locations, targeting precipitation (e.g., heavy rain), wind (e.g., hurricanes), or temperature (e.g., heat waves).
We observe that altering just 0.1\% of the measurements is sufficient to induce false extreme events and, consequently, trigger early warning systems in practice. This fraction is smaller than that corresponding to the input from a single polar-orbiting satellite. Nearly one hundred of these satellites are currently operated by different countries with partially conflicting political interests.
Furthermore, we demonstrate that an attacker can suppress actual extreme events, hindering timely preparations and potentially resulting in the loss of human lives. For example, we alter the predicted path of Hurricane Katrina (2005) to make it appear as though it would not strike New Orleans.

Our findings reveal a novel security threat that could erode trust in weather forecasting and have severe real-world consequences. As a potential defense, we investigate whether adversarial observations can be detected under theoretically ideal conditions. We find that detection success rates remain low (<3.1\%), indicating that detection-based strategies are unlikely to be effective in practice. Given that certifiably robust models are not yet available for weather forecasting, we argue that large-scale deployment of AI-based weather models should be delayed unless the underlying data sources can be fully trusted.

\vspace{-5pt}
\paragraph{Contributions.} In summary, we make the following major contributions in this work:
\begin{itemize}
    \setlength{\itemsep}{4pt}

\item \emph{Attack on weather forecasting.} We present the first attack targeting AI-based weather forecasting. Our attack is capable of creating adversarial observations that induce misleading predictions, such as non-existing extreme events, while remaining indistinguishable from natural noise.

\item \emph{Novel attack algorithm.} We propose a new algorithm for generating adversarial inputs for autoregressive diffusion models. The algorithm gradually approximates the inference process of weather prediction, achieving higher success rates than any existing attack.

\item \emph{Comprehensive evaluation.} We demonstrate the threat of adversarial observations by creating fake extreme events for a wide range of locations and time periods for the current best AI model GenCast. Additionally, we show that an attacker can suppress accurate extreme weather predictions.

\end{itemize}

To foster further research on the robustness of AI-based weather forecasting and to ensure the reproducibility of our experiments, we make our code and artifacts publicly available at 
\url{https://github.com/mlsec-group/adversarial-observations}. We also provide links to the considered weather datasets and models.

\paragraph{Roadmap.} We provide a brief introduction to weather forecasting in \cref{sec:background} before we present our attack in \cref{sec:method}. Our empirical analysis is provided in \cref{sec:eval}, and we investigate the detectability of the attack in \cref{sec:countermeasures}. We discuss the consequences of our findings and provide recommendations in \cref{sec:discussion}. Finally, we review related work in \cref{sec:related-work} and conclude in \cref{sec:conclusion}.

%% file: tex/02_background.tex
The goal of weather forecasting is to predict future weather conditions based on past observations. In this work, we focus on global weather forecasting, which is concerned with predicting weather patterns across the entire planet.
To this end, the global \emph{weather state} of the atmosphere, $\weatherstate \in \mathbb{R}^{|W|\times |V|}$, is represented as a grid of nodes $W$ distributed across the globe. Each node encodes a set of real-valued variables $V$ corresponding to key meteorological factors, such as temperature, wind speed, and sea level pressure.
By analyzing changes in the weather state over time, it becomes possible to estimate future conditions on the grid with varying degrees of confidence. Such forecasts underpin a wide range of practical applications, from predicting the output of solar and wind farms~\cite{solarWindForecast} to forecasting the paths of tropical cyclones~\cite{ullrich2021tempestextremes}.

Traditionally, weather forecasting has relied on \emph{numerical weather prediction}~(NWP) systems, which simulate the physical interactions between atmospheric variables to generate forecasts~\cite{ifs,ens}. These systems have long been the primary tool for global weather prediction. However, developing such models is highly resource-intensive and demands extensive domain expertise. Moreover, producing timely forecasts typically requires access to powerful supercomputers due to the substantial computational workload~\cite{natureNWPQuiet}.

\subsection{Learning-based Weather Prediction}
Machine learning-based weather prediction (MLWP) has recently emerged as an alternative to traditional forecasting. Rather than simulating physical processes explicitly, these models learn from historical weather data to infer atmospheric dynamics. This allows them to capture complex relationships between variables that reflect underlying physical laws. The latest MLWP systems outperform traditional methods in both accuracy and speed, producing high-quality forecasts in under ten minutes on a single computer~\mbox{\cite{gencast,graphcast,aardvark}}. Given their effectiveness, these models are highly attractive for practical use, and different efforts are underway to integrate them into operational weather forecasting~\cite{aifs, ai-noaa}.

\emph{GenCast}\cite{gencast} is currently the leading MLWP system, achieving the best performance~\cite{weatherbench2} in day-to-day forecasting as well as extreme event prediction. It employs an autoregressive diffusion model to generate sequential predictions of future weather states.
At its core is a denoising model $\denoisingStep$, which iteratively predicts the next state by denoising an initial estimate conditioned on the current and previous states of the global grid. This process is guided by the noise level of the initial estimate, which is gradually reduced over the denoising steps until the final prediction is obtained.

More formally, given the states $\weatherstate^{t-1}$ and $\weatherstate^t$, the model $\denoisingStep$ generates the next predicted state $\tilde{\weatherstate}^{t+1} = \mathbf{Z}_n^{t+1}$ by performing $n$ denoising steps. To this end, it begins with an initial sample $\mathbf{Z}^{t+1}_{0} \sim \mathcal{X}(\sigma_1)$ drawn from a noise distribution $\mathcal{X}$, parameterized by an initial noise level~$\sigma_1$.
Subsequently, each denoising step reduces the noise level from $\sigma_i$ to $\sigma_{i+1}$ according to the update rule,
\[
    \mathbf{Z}^{t+1}_{i+1} = \denoisingStep(\weatherstate^{t-1},
    \weatherstate^{t}, \mathbf{Z}^{t+1}_{i},\sigma_{i},\sigma_{i+1}).
\] 
where $\denoisingStep$ takes the past two states $\weatherstate^{t-1}$ and  $\weatherstate^{t}$, the current estimate $\mathbf{Z}^{t+1}_{i}$ as well as the respective noise levels $\sigma_{i}$ and $\sigma_{i+1}$ as input. This iterative and autoregressive refinement gradually enhances prediction detail by reducing noise at each step.

Training this model, however, poses a significant challenge: Backpropagating through all $n$ denoising steps is computationally prohibitive. As a remedy, the model is instead trained on individual denoising steps using:
\[
    \tilde{\weatherstate}^{t+1} = \denoisingStep(\weatherstate^{t-1}, \weatherstate^{t},  \mathbf{Z}, \sigma, 0) \quad \mathbf{Z} \sim \mathcal{X}(\sigma), \,\, \sigma \sim \Sigma(0, 1)\,,
\] 
where $\Sigma(a, b)$ is a probability distribution whose quantiles align with the noise levels $\sigma_1,\dots,\sigma_n$, spanning steps $a\cdot n$ to $b\cdot n$. During training, the full noise schedule with parameters $a = 0$ and $b = 1$ is used, thereby approximating the model's behavior across all $n$ diffusion steps.

Due to the random initialization of noise samples in each step, the prediction process is inherently stochastic. In the context of weather forecasting, this randomness is not necessarily a limitation. GenCast harnesses this stochasticity by generating multiple predictions, forming an ensemble that captures a range of plausible future scenarios. This ensemble-based approach enables uncertainty quantification and significantly enhances the so-called \emph{forecast skill}---the ability to make accurate predictions of the global weather state~\cite{gencast}. 
Interestingly, this randomness makes constructing effective perturbations more difficult than in deterministic models.

\subsection{Data Assimilation} 

Our discussion of weather forecasting still misses a key aspect: Real-world observations, such as temperature, pressure, and humidity, rarely align exactly with the points of a global grid. Instead, data from sources nearby the grid points must be integrated to form a consistent representation of the current weather state. This process, known as \emph{data assimilation} in meteorology, is essential for producing accurate forecasts.

\begin{figure}[bt!]
    \centering
    \includegraphics[width=0.9\columnwidth,height=0.4\columnwidth]{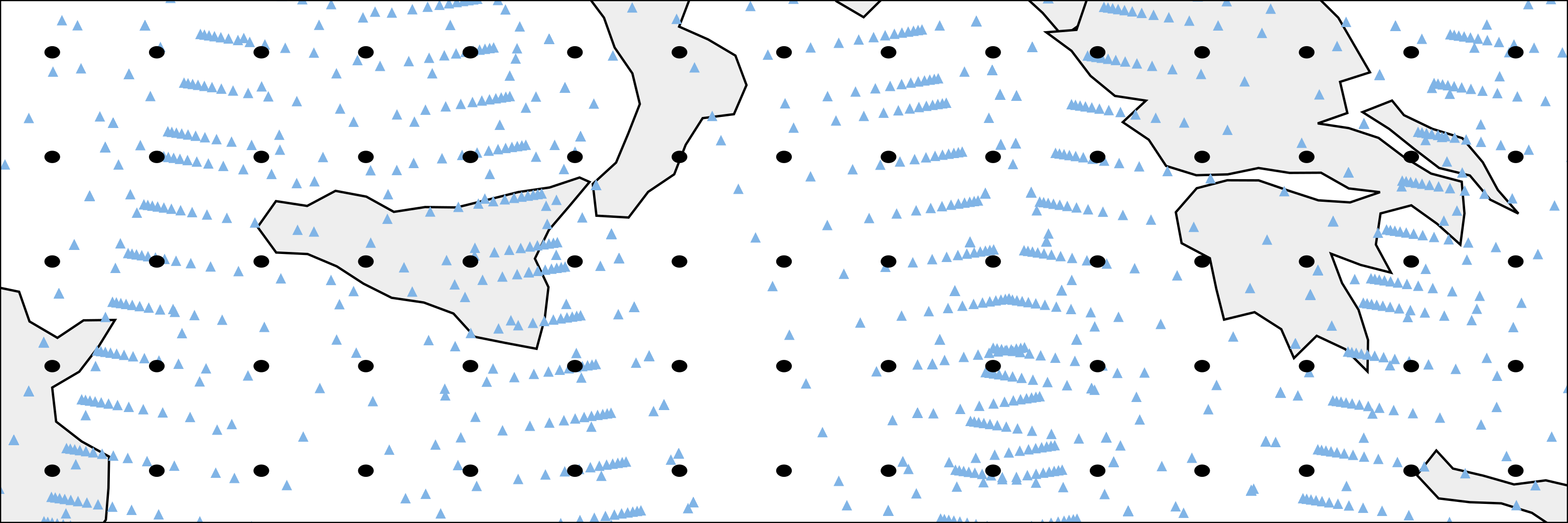}
    \caption{\textbf{Locations of satellite observations (blue~\raisebox{0.12ex}{\tikztriangle[poscolor]{1.5pt}}) and grid points (gray~\raisebox{0.12ex}{\tikzcircle[gridcolor,fill=gridcolor]{2pt}}) for a single prediction step.} 
    The satellite paths are computed based on the orbital elements of METOP-B, METOP-C and NOAA~15 as measured by NORAD~\cite{celetrak}.}
    \Description{The outline of the coast of southern Italy and Greece is seen with a regular grid of points placed across the whole image. Interspersed between the grid are triangles following straight paths. Groups of parallel paths have increasing distance but different groups are not necessarily parallel to each other.}
    \label{fig:grid}
\end{figure}

Data assimilation draws on a wide range of sources, from stationary observation points such as land stations and sea buoys to mobile platforms including balloons, aircraft, ships, and satellites~\cite{eyre2022assimilation}. Of these, satellites contribute by far the largest share, providing nearly 90\% of all assimilated data~\cite{pie-chart}. 
This dominance stems from the capability of satellites in polar orbit to scan the entire Earth’s surface approximately every 12 hours~\cite{iasi-metop} and geostationary satellites providing a near realtime view of a large area. Due to these capabilities, several international consortia operate meteorological satellites and contribute to global data assimilation, such as China's CMA and NRSCC with 3 satellites, the US NOAA, NASA and US Navy with 49 satellites, and Europe's EUMETSAT and ESA with 14 satellites. As an example, \Cref{fig:grid} shows measurements of three satellites within one prediction period and the respective grid points.

Technically, data assimilation involves making an initial estimate of the current atmospheric state at a grid point, then refining it through iterative optimization~\cite{le1986variational}. This process is driven by an objective that balances two main sources of error:
\begin{itemize}
\setlength{\itemsep}{4pt}
\item \emph{Observation error.} This error quantifies how closely the estimated state matches actual observations. For instance, if the observed surface temperature is 20\,°C but a nearby grid point predicts 0\,°C, the large discrepancy results in a high observation error.

\item \emph{Background error.} This error captures the deviation between the estimated state and a short-term forecast based on previously assimilated states. The short-term forecast incorporates past observational data, thus propagating historical information into the current estimate.
\end{itemize}

The assimilated state combines current observations with short-range forecasts derived from previous states, each of which carries inherent uncertainty. To account for this uncertainty, it is explicitly modeled within the data assimilation process. This typically involves estimating the noise through the variances of observation and background errors, which are then used to regularize the assimilation procedure~\cite{eyre2022assimilation}.
In contrast to the randomness in the GenCast model, this noise plays into the attacker’s hand, as it allows manipulations to be concealed within the expected uncertainty of the assimilated data as we show in the following.

%% file: tex/03_method.tex
Thus far, we have outlined how weather forecasting relies on observational data from numerous sources and is subject to inherent uncertainty in both data assimilation and inference. Building on this foundation, we now introduce our attack, which aims to manipulate forecasts by injecting adversarial observations. Before presenting the attack, we first describe the underlying threat model.

\subsection{Threat Model}\label{sec:threat-model}
We characterize the threat of adversarial observations in terms of the attacker’s goal, capabilities, and constraints. 

\paragraph{Attacker's goal}

We consider a scenario in which an attacker aims to manipulate forecasts generated by autoregressive diffusion models, such as those used in GenCast~\cite{gencast}. Potential attack goals range from causing economic harm by altering regional wind predictions, to inciting social disruption through fabricated extreme weather forecasts, and ultimately to causing physical harm by concealing impending disasters and preventing timely preparation.

\paragraph{Attacker's capabilities}
We assume that the adversary is capable of slightly manipulating the inputs to the forecasting model, specifically, the grid $\weatherstate$ assimilated from data of different meteorological sources (see \cref{sec:background}). While such manipulations could, in principle, be introduced at any source, we focus on measurements from polar-orbiting satellites due to their predominance in the assimilation process---contributing over 90\%~\cite{pie-chart}---and their ability to cover the entire Earth's surface within 12 hours.

Weather satellites are managed by meteorological and space agencies worldwide, including those operated by the USA, China, India, Germany, the European Union, Japan, France, and Taiwan~\cite{satellites}.
An attacker could compromise satellite data through various means, including internal sabotage, tampering with transmissions, breaches at ground-based command centers, or by exploiting vulnerabilities within the satellite systems~\cite{satellitesInsecure, hackingSattelite}. Even more concerning, manipulations could also be deliberately introduced by an operator as part of a strategic attack against another country.
Moreover, we assume that the adversary has white-box access to the forecasting model, including full knowledge of its architecture and parameters. In contrast to other domains, this assumption is plausible, as state-of-the-art learning models for weather forecasting are generally open-sourced~\cite[e.g.,][]{gencast,aardvark,aifs,graphcast}, as no significant security concerns have been raised so far.

\paragraph{Attacker's constraints} We assume that any manipulation of the forecasting model’s input is subject to practical constraints. For instance, control over a single satellite does not permit arbitrary modifications, as its observations are assimilated alongside data from numerous other sources. As a result, we assume that the adversary can modify only a small fraction of the values at each node in the weather state.
Note that polar-orbiting satellites pass over each grid point approximately twice per day, so global perturbations are surprisingly not a limiting factor in our attack.
In addition, manipulations are constrained by mechanisms designed to detect errors. Since weather forecasting is inherently imprecise, several such mechanisms are employed to reduce errors in the model’s input as early as possible. Consequently, manipulations are only effective if the introduced perturbations remain within the expected variance of the input variables. In this context, adversaries can exploit the noisy nature of weather measurements but cannot introduce larger deviations without risking detection.

To model these constraints, we assume that the adversary can introduce noise with a small standard deviation, denoted by $\epsilon$, where $\epsilon$ is smaller than the expected variance of any variable at the manipulated grid points. Furthermore, we conservatively assume that the perturbation must be unbiased, as the adversary can influence only a limited portion of the collected observational data.

\subsection{Attack Methodology}\label{sec:attack}

Building on our threat model, we now present our attack strategy for generating adversarial observations. The core idea is to manipulate the estimated state $\tilde{\weatherstate}^{t}$ at time $t$ so that the predicted state $\tilde{\weatherstate}^{t+j}$ at a later time $t + j$ aligns with a predefined target. 
To achieve this, the attack adds perturbations $\perturbation^t$ and $\perturbation^{t-1}$ to the observed states $\weatherstate^t$ and $\weatherstate^{t-1}$, respectively, thereby influencing the calculation of $\tilde{\weatherstate}^{t+j}$ in the subsequent  autoregressive iterations. The perturbations are constrained to be unbiased and limited in magnitude, with standard deviations not exceeding a threshold $\epsilon$.

\paragraph{Objective function}
Formally, the objective can be defined through an adversarial loss function $\advloss$, which measures the distance from a selected  target and is minimized by the attacker using the (approximated) inference function $\inferenceFunction$ of the MLWP system. In the case of GenCast, this function encapsulates the entire prediction procedure across multiple noise levels and time steps.

To model the constrained perturbations, we define a per-variable mean $\mu_v$ and standard deviation $\sigma_v$ for each variable $v \in V$, which the perturbations must satisfy. These parameters allow us to constrain both the direction and the variability of the adversarial influence. Combining these elements, we arrive at the following optimization problem:
\begin{align*}
    \argmin_{\perturbation^t,\perturbation^{t-1}}& \quad \advloss\left(\inferenceFunction(\weatherstate^{t}+\perturbation^{t},\weatherstate^{t-1}+\perturbation^{t-1},j,n)\right) \\
    \text{subject to}&\quad \forall v\in V: \mu_v = 0 \land \sigma_v \leq \epsilon,
\end{align*}
where $j$ is the lead time for the forecast and $n$ the number of considered noise levels used within $\inferenceFunction$.

\paragraph{Decomposing the adversarial loss} The function $\advloss$ captures the complex task of manipulating forecasts in a single expression, rendering direct optimization challenging. To address this, we decompose $\advloss$ into two modular components, $\mathcal{A} = \mathcal{V} \circ \mathcal{S}$. The spatial function $\mathcal{S}$ specifies the geographic region of interest, while the variable function $\mathcal{V}$ extracts the relevant meteorological target within that region. This structured formulation provides a flexible and unified optimization framework, capable of representing a wide range of targets—from fabricating extreme winds to concealing genuine rainfall anywhere on the globe.

To illustrate the utility of this decomposition, let us consider the following definition of the adversarial loss $\advloss$:
\begin{align*}
    \mathcal{S} &\colon \weatherstate \mapsto \left\{\weatherstate_{(\text{lat},\text{lon})} \mid \text{lat}\in [51, 52], \text{lon}\in [-1, 1]\right\}, \\
    \mathcal{V} &\colon R \mapsto -\min_{r\in R}\left(\sqrt{(r_\text{u-wind})^2 + (r_\text{v-wind})^2}\right). 
\end{align*}
In this example, the spatial function $\mathcal{S}$ selects all grid points with latitudes between $51^\circ$ and $52^\circ$ and longitudes between $-1^\circ$ and $1^\circ$, corresponding to the London area. The variable function $\mathcal{V}$ then computes a scalar value from this region---specifically, negative minimum wind speed, derived from the eastward (U) and northward (V) wind components. As a result, the formulated loss function seeks to maximize the minimum predicted wind speed around London. More complex objectives can similarly be defined by customizing the spatial and variable functions.

\paragraph{Approximating the inference function}
The diffusion model underlying $\inferenceFunction$ is inherently non-deterministic, as it generates forecasts by iteratively denoising samples initialized with random noise. In the case of GenCast, this process unfolds over 40 steps, making end-to-end differentiation computationally prohibitive. To mitigate this, we could adopt an approximation strategy proposed by~\citet{advdm}, in which a single noise level is selected and the sample is denoised from that point onward.

However, this approximation alone does not fully resolve the challenge of determining effective perturbations for $\inferenceFunction$. First, the stochastic nature of the diffusion process means that the impact of a perturbation heavily depends on the specific realization of the initial noise. Second, the influence of the denoising step varies with the selected noise level: lower noise levels result in only minor forecast changes, while higher noise levels permit more substantial alterations. As a result, the optimization process becomes highly variable, and sampling only a single the noise level, as proposed by~\citet{advdm}, does not yield reliable perturbations for executing an attack.

\paragraph{Sampling multiple noise levels}
To improve the approximation of the inference process, we introduce two key refinements, as outlined in \Cref{alg:better_approx}. First, rather than selecting a single noise level, we sample $n > 1$ distinct levels drawn from non-overlapping intervals across the noise distribution. Second, instead of denoising in a single step, we perform a sequence of denoising operations: the process begins with noise sampled at the first level, followed by iterative denoising through the subsequent levels, and concludes with a final denoising step from the last level to zero.

\begin{algorithm}[h]
    \caption{Our approximation of the autoregressive diffusion inference process. }\label{alg:better_approx}
    \KwIn{inputs $\weatherstate^{t},\weatherstate^{t-1}$, lead time steps $j$, number of steps $n$}
    \KwOut{approximate prediction $\tilde{\weatherstate}^{t+j}$}
    $\mathbf{Z}^{t}_{n}, \mathbf{Z}^{t-1}_{n} \gets \weatherstate^{t},\weatherstate^{t-1}$\;
    \For{$\tau \gets t+1$ \KwTo $t+j$}{
        Sample $\sigma_0,\dots,\sigma_{n-1} \sim \Sigma\left(0,\frac{1}{n}\right),\dots,\Sigma\left(\frac{n-1}{n},1\right)$\;
        Sample $\mathbf{Z}^{\tau}_{0} \sim \mathcal{X}(\sigma_0)$\;

        \For{$i \gets 1$ \KwTo $n - 1$}{
            $\mathbf{Z}^{\tau}_{i} \gets \denoisingStep(\mathbf{Z}^{\tau - 2}_{n}, \mathbf{Z}^{\tau -1}_{n},\mathbf{Z}^{\tau}_{i-1},\sigma_{i-1},\sigma_{i})$\;
        }
        $\mathbf{Z}^\tau_{n} \gets \denoisingStep(\mathbf{Z}^{\tau - 2}_{n}, \mathbf{Z}^{\tau -1}_{n},\mathbf{Z}^{\tau}_{n-1},\sigma_{n-1},0)$\;
    }
    
    \Return{$\mathbf{Z}^{t+j}_{n}$}\;
\end{algorithm}

This strategy ensures that each optimization step incorporates both high and low noise levels, striking a balance between influence and difficulty. In doing so, our refined approximation more closely mimics the full inference procedure, which spans the entire range of noise levels.
In particular, lines 3--4 of \Cref{alg:better_approx} sample from the aligned distribution $\Sigma$ and the noise distribution $\mathcal{X}$ to generate an initial estimate. Subsequently, lines 5--7 iteratively refine this estimate by applying the denoising function $\denoisingStep$ across a sequence of decreasing noise levels $\sigma_i$.

\paragraph{Projecting the perturbations}
Finally, to ensure that the perturbations remain within the prescribed bounds, we introduce a projection operator $\Pi$, defined as
\[
    \Pi_\epsilon(\perturbation) = (\perturbation - \mu_{v}) \cdot \frac{\min(\epsilon,\sigma_{v})}{\sigma_{v}}.
\] 
This operator is applied to the perturbation $\perturbation$ of each variable $v$ across all grid points during optimization. We denote this as $\Pi_\epsilon(\perturbation)$, indicating that the projection is performed independently for each variable. The projection ensures that the perturbations conform to the specified per-variable constraints, maintaining the prescribed mean $\mu_v$ and standard deviation $\sigma_v$.

\paragraph{Complete attack algorithm}
The complete attack procedure, integrating all components and refinements, is presented in \Cref{alg:attack}. The method follows a standard gradient-based framework for generating adversarial inputs with $N$ iterations, leveraging the approximated inference function (line 4) and applying the projection operator $\Pi$ to enforce perturbation constraints (lines 6 and 9). To improve optimization efficiency, we incorporate momentum into the gradient updates (line 6) and use a cosine annealing schedule (line 7) to dynamically adjust the step size throughout the process.

\begin{algorithm}[h]
    \caption{Our attack algorithm with $n$ approximation steps of the diffusion process.}\label{alg:attack}
    \KwIn{attack budget \(\epsilon\), number of attack steps $N$, lead time steps $j$, inputs $\weatherstate^{t},\weatherstate^{t-1}$}
    \KwOut{adversarial perturbation $\perturbation^{t},\perturbation^{t-1}$}

    $\mathbf{m}_0 \gets \mathbf{0}$\;
    $\perturbation_0 = (\perturbation^{t}_0, \perturbation^{t-1}_0) \gets \mathbf{0}$\;

    \For{$i \gets 1$ \KwTo $N$}{
        $\tilde{\weatherstate}^{t+j} = \inferenceFunction(\weatherstate^{t}+\perturbation^t_{i-1},\weatherstate^{t-1}+\perturbation^{t-1}_{i-1},j,n)$\;
        $\mathbf{g}_i \gets \nabla_{\perturbation_{i-1}}
            \advloss (\tilde{\weatherstate}^{t+j})$\;
        $\mathbf{m}_i \gets
        \beta \cdot \mathbf{m}_{i-1} +
        (1 - \beta) \cdot \Pi_1\left(\mathbf{g}_i
        \right)$\;
        $\alpha'_i \gets \frac{\epsilon}{N} + \frac{1}{2} \left(2 \epsilon - \frac{\epsilon}{N}\right) \cdot \left(1 + \cos\left(\frac{(i - 1) \cdot \pi}{N}\right)\right)$\;
        $\alpha_i \gets \frac{\alpha'_i}{(1 - \beta)^i}$\;
        $\perturbation_i \gets \Pi_\epsilon (\perturbation_{i-1} - \alpha_i \mathbf{m}_i)$;
    }
    \Return{$\perturbation^{t}_N,\perturbation^{t-1}_N$}
\end{algorithm}

%% file: tex/04_evaluation.tex
\begin{figure*}[tb!]
    \centering
    \input{tikz/main_eval}
    \vspace{-0.5cm}
    \caption{\textbf{Resulting mean deviation induced by adversarial observations of different sizes.} The average deviation of wind speed, temperature and precipitation as well as the 90\% confidence interval across all target locations and times are shown. The attacker goal is to achieve the threshold for 99\% extreme weather deviations with minimal noise increase.}
    \Description{Three figures show the average deviation in prediction induced by different attack methods at different attack budgets, represented by the increase in total noise. Each figure shows the effect on a single target variable. Common between all figures is that our attack induces a larger deviation at each attack budget than the baseline attacks. Furthermore, a far smaller increase in noise is needed to pass a deviation that would qualify as extreme weather. The trends are similar for wind speed and temperature, but exacerbated for precipitation, where the gap between our attack and the baselines is even larger.}
    \label{fig:main-eval}
\end{figure*}
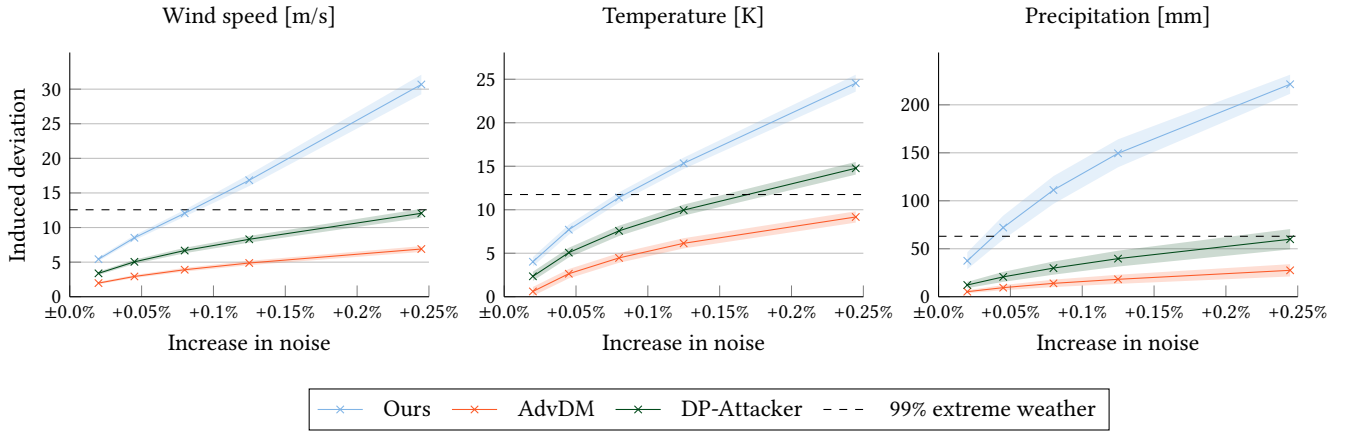

We proceed to evaluate the effectiveness of the proposed attack in generating adversarial observations under real-world conditions. To this end, we consider two scenarios: (a) fabricating extreme events and (b) concealing extreme events. That is, we first investigate whether adversarial observations can reliably induce non-existent extreme events across various locations and points in time. Second, we examine whether the accuracy of forecasts for genuine extreme events can be compromised, for example, by moving their location or diminishing their intensity.

\subsection{Experimental Setup} 
\label{sec:setup}
For all our experiments, we target GenCast~\cite{gencast}, the currently leading MLWP system~\cite{weatherbench2}. Specifically, we use the median prediction deviation from a GenCast ensemble consisting of five members and consider a one-degree grid resolution. For our attack, we generate adversarial observations two days prior to a target prediction time with $j=4$, resulting in an attack time offset of two days. We use $N = 50$ iterative optimization steps to ensure that the resulting deviation remains robust to the stochasticity of the inference process. Additionally, we set the number of approximation steps per iteration to $n = 2$. All experiments were run on server with four NVIDIA A40 GPUs.

\paragraph{Dataset}
We perform all experiments using the ERA5 dataset~\cite{era5}, which provides hourly assimilated weather variables across multiple pressure levels and covers the entire globe. This is the same dataset on which GenCast was trained.
For a single state $\weatherstate$ this amounts to approximately $>5\,\text{M}$ individual variable values, distributed across 65,160 grid points.
We evaluate on data from 2022, which is the most recent full year that is publicly available as part of WeatherBench2~\cite{weatherbench2}, the most common benchmark for MLWP systems. 

\paragraph{Extreme weather}
Following common practice in meteorology, we define extreme weather events based on the deviation of a target variable from its expected value. Specifically, we consider events exceeding the \emph{99th percentile} for three variables: (a) wind speed at 10 meters above ground, (b) temperature at 2 meters above ground, and (c) precipitation accumulated over a 12-hour period. That is, we focus on wind speed, temperature, and precipitation values in the top $1\,\%$ of measurements at each location.

To determine the 99th percentile threshold for each variable, we analyze all historical weather states available in the ERA5 dataset, evaluating each target variable individually. We construct a climatological model for each variable, estimating the expected value for any given day of the year at a specific location by averaging across all available years. Using this model, we compute the maximum deviation between the expected and actual values of the variable for each year and location. We then derive the 99th percentile of these yearly maxima and average them across all grid points to obtain thresholds corresponding to the 99\% extreme weather deviations.

\paragraph{Attacker setup}
For our attack, we assume an adversary capable of manipulating data from a single polar-orbiting satellite. Under this scenario, we derive the maximum permissible standard deviation~$\epsilon$ (see \Cref{sec:threat-model}). Since the individual contribution of a single satellite cannot be precisely determined, we conservatively approximate its influence by assuming it is smaller than average:
Approximately 100 meteorological satellites contribute to the ECMWF assimilation system~\cite{satellites}, so that, on average, a single satellite accounts for more than 1\% of the total observation error.
Since this error is typically larger than background error, we can set a lower bound on it using the background error~\cite{obsErrors}.
Specifically, we limit the increase in noise to just 0.25\,\% of the standard deviation of the background error.

To map this relative constraint to absolute terms, we estimate the variance of the background error per year. As previously described, the background error is defined as the difference between the short-range forecast from the previous state and the final assimilated state. We use GenCast to perform a single-step forecast for each of our evaluation years and compute the difference to the corresponding assimilated values. Finally, we calculate the average variance across all grid points and forecasts for each variable.
The resulting attack setup is conservative and clearly underestimates the potential real-world impact of compromising a single satellite.

\subsection{Fabricating Extreme Events}

We begin by investigating whether adversarial observations can trigger extreme weather predictions across different locations and times.
To select target locations, we focus on densely populated areas. Specifically, we randomly sample 100 sites from the 1,000 most populous population centers using the Global Human Settlement Urban Centre Database R2024A~\cite{ghs:ucdb}, which provides up-to-date estimates of global population distribution based on satellite imagery analysis. The selected locations range from mid-sized cities such as Suez and Leipzig to major metropolitan areas like Los Angeles and Ho Chi Minh City. For each site, we randomly select a target time within the evaluation year 2022.

For each of these location–time pairs, we run our attack to induce extreme deviations in each of the three target weather variables at the specified location and time, manipulating the observations two days earlier $(j = 4)$. To evaluate the impact of perturbation strength, we conduct the attack using logarithmically spaced noise budgets, starting from $0.02\,\%$ and increasing up to the derived maximum of $0.25\,\%$, as discussed in \Cref{sec:setup}.

\paragraph{Attack performance}
The results of this experiment are shown in \Cref{fig:main-eval}, displaying the deviations across all target variables.
We observe that adversarial observations consistently induce substantial changes in weather predictions. For each of the three target variables, the noise required to exceed the deviation threshold remains well below the maximum allowed perturbation of $0.25\,\%$.
On average, triggering extreme weather conditions for temperature and wind speed requires a noise level of approximately $0.08\,\%$, while precipitation proves even more sensitive, with the threshold for extreme weather surpassed at noise levels below $0.05\,\%$.

To put these numbers into perspective, at the maximum permitted noise level of $0.25\,\%$, the attack can increase wind speeds by $30.7\,\text{m/s}$---equivalent to $111\,\text{km/h}$---\emph{averaged} over a 12-hour period. This average is on par with peak wind speeds typically observed during a Category 1 hurricane. 
Similarly, temperatures can be increased by $24.6\,\text{C}$, while precipitation can be increased by $221\,\text{mm}$ over a 12-hour period---equivalent to $221\,\text{l/m}^2$. This level of rainfall is comparable to that seen during extreme storm events.
These results demonstrate that even minimal perturbations to observations can lead to substantial shifts in forecast outputs, highlighting the vulnerability of state-of-the-art weather prediction systems to adversarial manipulation.

\paragraph{Baselines}
Next, we consider the performance of our attack against two recently proposed methods targeting diffusion models. The first, AdvDM~\citep{advdm}, introduces perturbations directly into the noise used by image diffusion models.
The second, DP-Attacker~\citep{dp-attacker}, is a more recent method designed to target policy diffusion models. These models generate multi-step policies autoregressively from an initial vision input, which is more similar to weather prediction and makes this attack naturally suited to our context.
Both baseline attacks operate using a single sampled noise level for prediction, consistent with the noise sampling employed during training.

The results are included in \Cref{fig:main-eval}. Our attack consistently outperforms the baseline methods across all target variables and attack budgets. Notably, both baselines fail to reach the extreme weather thresholds for wind speed and precipitation. Only DP-Attacker achieves the temperature threshold with an attack budget below the maximum noise increase of $0.25\,\%$.
When comparing the baselines with our method, we observe that the performance gap widens as the attack budget increases. This suggests that our approach scales more efficiently as the budget grows. This advantage is particularly evident in the case of precipitation, where our method surpasses the baselines by a substantially larger margin.

\begin{figure}[t]
    \centering
    \input{tikz/susceptibility}
    \caption{\textbf{Mean required noise increase at different locations.} The dashed line shows a linear regression of the required noise. The mean increase in noise required to fabricate an extreme weather prediction grows with increasing distance from the equator.}
    \Description{A plot depicting scattered points and a trend line. The y axis ranges from zero percent to 0.25 percent and shows the mean increase in noise needed to fabricate an extreme weather event at a particular location. The x axis shows the absolute latitude, ranging from 0 to 60. The scattered points generally follow the trend line that is positively inclined, indicating more noise is needed for locations further from the equator. Some outliers do not follow this trend and instead need much more noise.}
    \label{fig:susceptibility}
\end{figure}
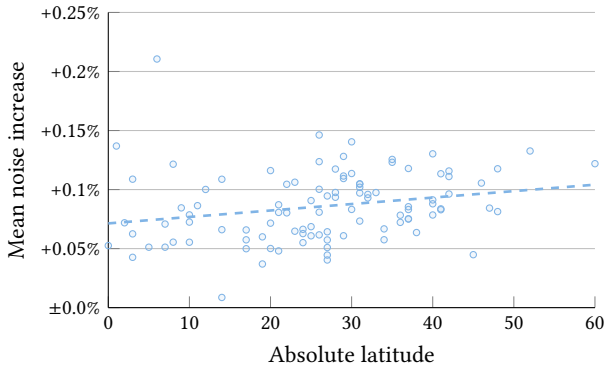

\begin{table}[b]
    \caption{\textbf{Mean relative deviation achieved by different ablations.} The deviation is relative to the original attack and averaged across $200$ different target combinations.}
    \label{tab:ablation}
    \centering
    \input{tables/ablation}
\end{table}

\paragraph{Susceptibility of different locations}

To explore how the choice of target location influences the attack, we investigate whether predictions at certain locations are more susceptible to adversarial observations than others. This is evaluated by calculating the average increase in noise required at each target location to achieve extreme weather, averaged over all target variables. We estimate this by linearly interpolating the induced deviations between the observed values.

Our findings, illustrated in \Cref{fig:susceptibility}, indicate a relationship between the required noise and the angular distance from the equator. Specifically, locations farther from the equator tend to require more noise to achieve the same level of deviation ($p < 0.05$).  Still, even the most impacted areas require less than the maximum possible noise increase to trigger an extreme weather prediction---indicating that, although the effect is statistically meaningful, its practical impact is relatively modest.
We hypothesize that this trend is linked to the uneven distribution of grid points near the equator. Because the grid is constructed with uniform spacing in both latitude and longitude, grid points become increasingly dense toward the poles and more sparse near the equator. Near-equatorial cells can span over $100\,km$ (approximately $70\,\text{miles}$) per side.
To address this imbalance, one potential solution is to use a mesh derived from an icosahedron for input to the MLWP, which ensures uniform spacing between grid points regardless of geographic location. This approach aligns well with existing infrastructure, as GenCast already employs a six-times refined icosahedral grid internally. However, this adjustment alone does not resolve the underlying vulnerability.

\paragraph{Ablation study}

To better understand the contribution of individual components within our attack methodology, we perform an ablation study. Specifically, we evaluate three simplified variants: (1) replacing our improved approximation of the inference process with the naïve approach used during training, (2) removing optimization enhancements such as cosine annealing and momentum, and (3) removing both steps simultaneously.
Due to computational constraints, we restrict our evaluation to a subset of 200 out of the original 1,500 target combinations. For each variant, we compute the relative deviation in performance compared to the full attack, quantifying the extent to which each component contributes to overall effectiveness.

The average relative deviations for the three considered variants across different target variables are shown in \Cref{tab:ablation}. Removing any single component lead to a noticeable drop in performance.
For temperature and wind speed, most of the performance is retained when only the improved optimization steps are removed. This suggests that the inference approximation plays a more critical role for these variables. Moreover, removing the inference approximation alone has a larger impact than removing the improved steps.
As expected, disabling both components results in the largest performance reduction. These findings suggest that the interplay between the inference approximation and the optimization enhancements is essential to achieving the strong attack effectiveness observed in our earlier experiments.

%% file: tikz/main_eval.tex
\begin{tikzpicture}
\pgfplotsset{every tick label/.append style={font=\small}}
\begin{groupplot}[
    group style={
        group size= 3 by 1,
        horizontal sep=1.0cm,
    },
    xlabel={Increase in noise},
    ylabel={},
    axis x line*=bottom,
    axis y line*=left,
    xmin=0.0,
    xmax=0.0025,
    width=0.75\columnwidth,
    height=0.57\columnwidth,
    ymajorgrids,
    legend cell align={left},
    xtick={0,0.0005,0.001,0.0015,0.002,0.0025},
    xticklabels={$\pm$0.0\%,+0.05\%,+0.1\%,+0.15\%,+0.2\%,+0.25\%},
    xtick scale label code/.code={},
]
    \nextgroupplot[
        title={Wind speed [m/s]},
        ylabel={Induced deviation},
        ymin=0,
        ytick={0,5,10,15,20,25,30},
        yticklabels={0,5,10,15,20,25,30},
    ]
    \addplot [mark=x, poscolor] table [x index=0,y index=1] {data/main_eval/wind/ours.dat};
    \addplot [name path=lolower, fill=none, draw=none, forget plot] table [
       x index=0,
       y expr=\thisrowno{1} - \thisrowno{2}]{data/main_eval/wind/ours.dat};
    \addplot [name path=loupper, fill=none, draw=none, forget plot] table [
       x index=0,
       y expr=\thisrowno{1} + \thisrowno{2}]{data/main_eval/wind/ours.dat};
    \addplot[poscolor, opacity=0.2, forget plot] fill between[of=lolower and loupper];

    \addplot [mark=x, negcolor] table [x index=0,y index=1] {data/main_eval/wind/advdm.dat};
    \addplot [name path=lolower, fill=none, draw=none, forget plot] table [
       x index=0,
       y expr=\thisrowno{1} - \thisrowno{2}]{data/main_eval/wind/advdm.dat};
    \addplot [name path=loupper, fill=none, draw=none, forget plot] table [
       x index=0,
       y expr=\thisrowno{1} + \thisrowno{2}]{data/main_eval/wind/advdm.dat};
    \addplot[negcolor, opacity=0.2, forget plot] fill between[of=lolower and loupper];

    \addplot [mark=x, neucolor] table [x index=0,y index=1] {data/main_eval/wind/dp-attacker.dat};
    \addplot [name path=lolower, fill=none, draw=none, forget plot] table [
       x index=0,
       y expr=\thisrowno{1} - \thisrowno{2}]{data/main_eval/wind/dp-attacker.dat};
    \addplot [name path=loupper, fill=none, draw=none, forget plot] table [
       x index=0,
       y expr=\thisrowno{1} + \thisrowno{2}]{data/main_eval/wind/dp-attacker.dat};
    \addplot[neucolor, opacity=0.2, forget plot] fill between[of=lolower and loupper];
    
    \addplot[mark=none, black, dashed, samples=2, domain=0:0.0025] {12.57};

    \nextgroupplot[
        title={Temperature [K]},
        ymin=0,
        ytick={0,5,10,15,20,25},
        yticklabels={0,5,10,15,20,25},
    ]
    \addplot [mark=x, poscolor] table [x index=0,y index=1] {data/main_eval/temperature/ours.dat};
    \addplot [name path=lolower, fill=none, draw=none, forget plot] table [
       x index=0,
       y expr=\thisrowno{1} - \thisrowno{2}]{data/main_eval/temperature/ours.dat};
    \addplot [name path=loupper, fill=none, draw=none, forget plot] table [
       x index=0,
       y expr=\thisrowno{1} + \thisrowno{2}]{data/main_eval/temperature/ours.dat};
    \addplot[poscolor, opacity=0.2, forget plot] fill between[of=lolower and loupper];

    \addplot [mark=x, negcolor] table [x index=0,y index=1] {data/main_eval/temperature/advdm.dat};
    \addplot [name path=lolower, fill=none, draw=none, forget plot] table [
       x index=0,
       y expr=\thisrowno{1} - \thisrowno{2}]{data/main_eval/temperature/advdm.dat};
    \addplot [name path=loupper, fill=none, draw=none, forget plot] table [
       x index=0,
       y expr=\thisrowno{1} + \thisrowno{2}]{data/main_eval/temperature/advdm.dat};
    \addplot[negcolor, opacity=0.2, forget plot] fill between[of=lolower and loupper];

    \addplot [mark=x, neucolor] table [x index=0,y index=1] {data/main_eval/temperature/dp-attacker.dat};
    \addplot [name path=lolower, fill=none, draw=none, forget plot] table [
       x index=0,
       y expr=\thisrowno{1} - \thisrowno{2}]{data/main_eval/temperature/dp-attacker.dat};
    \addplot [name path=loupper, fill=none, draw=none, forget plot] table [
       x index=0,
       y expr=\thisrowno{1} + \thisrowno{2}]{data/main_eval/temperature/dp-attacker.dat};
    \addplot[neucolor, opacity=0.2, forget plot] fill between[of=lolower and loupper];
    
    \addplot[mark=none, black, dashed, samples=2, domain=0:0.0025] {11.75};

    \nextgroupplot[
        title={Precipitation [mm]},
        ymin=0,
        ytick={0,0.05,0.1,0.15,0.2},
        yticklabels={0,50,100,150,200},
        legend style={legend columns=4,column sep=0.2cm},legend to name={commonLegend},
    ]
    \addplot [mark=x, poscolor] table [x index=0,y index=1] {data/main_eval/precipitation/ours.dat};
    \addlegendentry{Ours}
    \addplot [name path=lolower, fill=none, draw=none, forget plot] table [
       x index=0,
       y expr=\thisrowno{1} - \thisrowno{2}]{data/main_eval/precipitation/ours.dat};
    \addplot [name path=loupper, fill=none, draw=none, forget plot] table [
       x index=0,
       y expr=\thisrowno{1} + \thisrowno{2}]{data/main_eval/precipitation/ours.dat};
    \addplot[poscolor, opacity=0.2, forget plot] fill between[of=lolower and loupper];

    \addplot [mark=x, negcolor] table [x index=0,y index=1] {data/main_eval/precipitation/advdm.dat};
    \addlegendentry{AdvDM}
    \addplot [name path=lolower, fill=none, draw=none, forget plot] table [
       x index=0,
       y expr=\thisrowno{1} - \thisrowno{2}]{data/main_eval/precipitation/advdm.dat};
    \addplot [name path=loupper, fill=none, draw=none, forget plot] table [
       x index=0,
       y expr=\thisrowno{1} + \thisrowno{2}]{data/main_eval/precipitation/advdm.dat};
    \addplot[negcolor, opacity=0.2, forget plot] fill between[of=lolower and loupper];

    \addplot [mark=x, neucolor] table [x index=0,y index=1] {data/main_eval/precipitation/dp-attacker.dat};
    \addlegendentry{DP-Attacker}
    \addplot [name path=lolower, fill=none, draw=none, forget plot] table [
       x index=0,
       y expr=\thisrowno{1} - \thisrowno{2}]{data/main_eval/precipitation/dp-attacker.dat};
    \addplot [name path=loupper, fill=none, draw=none, forget plot] table [
       x index=0,
       y expr=\thisrowno{1} + \thisrowno{2}]{data/main_eval/precipitation/dp-attacker.dat};
    \addplot[neucolor, opacity=0.2, forget plot] fill between[of=lolower and loupper];

    \addplot[mark=none, black, dashed, samples=2, domain=0:0.0025] {0.063};
    \addlegendentry{99\% extreme weather}

\end{groupplot}
    \node at (8.5, -1.5) {\pgfplotslegendfromname{commonLegend}};
\end{tikzpicture}

%% file: tikz/susceptibility.tex
\begin{tikzpicture}
    \pgfplotsset{every tick label/.append style={font=\small}}
    \begin{axis}[
        title={},
        ylabel={Mean noise increase},
        xlabel={Absolute latitude},
        ymin=0,
        ymax=0.0025,
        ytick={0,0.0005,0.001,0.0015,0.002,0.0025},
        yticklabels={$\pm$0.0\%,+0.05\%,+0.1\%,+0.15\%,+0.2\%,+0.25\%},
        ytick scale label code/.code={},
        axis x line*=bottom,
        axis y line*=left,
        xmin=0,
        xmax=60,
        width=0.95\columnwidth,
        height=0.65\columnwidth,
        ymajorgrids,
    ]
        \addplot [mark size=1.2,poscolor,fill opacity=0.1,only marks] table [x index=0,y index=1] {data/location/lat_susceptibility.dat};
        \addplot[mark=none, poscolor,dashed,line width=1pt, samples=2, domain=0:60] {5.466218016266582e-06*x + 0.0007134301027012592};
    \end{axis}
\end{tikzpicture}

%% file: tables/ablation.tex
\begin{tabular}{l rrr}
    \toprule
    \textbf{Method} & \multicolumn{1}{c}{\textbf{Wind Speed}} & \multicolumn{1}{c}{\textbf{Temperature}} & \textbf{Precipitation} \\
    \midrule
    Ours       & $100.0\,\%$ & $100.0\,\%$ & $100.0\,\%$ \\
    w/o steps  & $89.3\,\%$ ~(\red{-10.7}) & $93.1\,\%$ ~(\red{-~6.9}) & $54.4\,\% ~(\red{-45.6})$ \\
    w/o approx & $59.3\,\%$ ~(\red{-40.7}) & $71.6\,\%$ ~(\red{-28.4}) & $33.9\,\% ~(\red{-66.1})$ \\
    w/o both   & $56.0\,\%$ ~(\red{-44.0}) & $62.9\,\%$ ~(\red{-37.1}) & $18.4\,\% ~(\red{-81.6})$ \\
    \bottomrule
\end{tabular}

%% file: tex/05_case_study.tex
Thus far, our analysis has focused on scenarios in which an adversary seeks to fabricate predictions of extreme weather at specific times and locations. We now turn to a different question: can the attack also undermine genuine forecasts of extreme weather events? To explore this, we apply our method to three major historical events---Cyclone Amphan (2020), the 2006 European heat wave, and Hurricane Katrina (2005).
For each event, we simulate an attack by introducing a maximum noise perturbation of $\epsilon \leq 0.25\,\%$ into the weather predictions two and a half days before each event reached peak intensity. This time frame ensures that the extent and location of the extreme event is predicted correctly without the attack but could still be realistically manipulated.
Differing from the previous section, the attacker's goal is not to force extreme predictions at a single location on the grid but instead reducing the estimated intensity in an entire region. 

\begin{figure}[b]
    \centering
    \begin{subfigure}[t]{0.4\columnwidth}
        \centering
        \includegraphics[width=0.95\linewidth,height=1.3\linewidth]{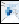}
        \caption{Original prediction}\label{fig:rain-unperturbed}
    \end{subfigure}
    \begin{subfigure}[t]{0.4\columnwidth}
        \centering
        \includegraphics[width=0.95\linewidth,height=1.3\linewidth]{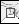}
        \caption{Perturbed prediction}\label{fig:rain-perturbed}
    \end{subfigure}
    \begin{subfigure}[t]{0.1\columnwidth}
        \centering
        \includegraphics[width=0.95\linewidth,height=5.2\linewidth]{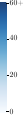}
    \end{subfigure}
    \caption{\textbf{Predicted precipitation at the peak of Cyclone Amphan.} The forecast is shown $($a$)$ without attack and $($b$)$ after including adversarial observations. The dashed rectangle depicts the target region of the attack. The precipitation is expressed as mm over a 12-hour period.}
    \Description{Two images both showing the outline of South Asia, in particular Bangladesh and parts of India. The image on the left, that is labeled original prediction shows a region of blue and dark blue rectangles in the middle, indicating heavy rainfall, as per the colorbar on the right. The image on the right, containing a dashed rectangle around the previously rainy region, is devoid of any blue rectangle showing rain.}
    \label{fig:rain-case-study}
\end{figure}

\paragraph{Cyclone Amphan}
Our first case study focuses on tropical Cyclone Amphan, which struck Bangladesh, India, and Sri Lanka in May 2020, bringing strong winds and heavy rainfall that caused widespread flooding~\cite{Kumar2021}.
Several days prior to landfall, the storm significantly intensified---which was correctly predicted by GenCast---leading up to intense precipitation across the region as shown in \Cref{fig:rain-unperturbed} as the blue shaded area.

We adversarially perturb the observations before this intensification, targeting a prediction outcome with minimal precipitation across the expected storm region. As shown in \Cref{fig:rain-perturbed}, the resulting forecast entirely suppresses precipitation in the target region.
Notably, when examining the sequence of predicted states between the perturbed inputs and the forecast, we observe a plausible dissipation of the storm. In this manipulated scenario, the storm releases rainfall over the ocean and weakens before reaching land. This illustrates how an attacker could convincingly mask an otherwise accurate forecast of a severe weather event. Crucially, the perturbations as well as the intermediate weather development appear plausible despite the underlying manipulations.

\begin{figure}[t]
    \centering
    \includegraphics[width=0.8\columnwidth]{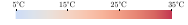}
    \begin{subfigure}{\columnwidth}
        \centering 
        \includegraphics[width=0.855\columnwidth,height=0.54\columnwidth]{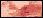}
        \caption{Original prediction}\vspace{0.25cm}
    \end{subfigure}
    \begin{subfigure}{\columnwidth}
        \centering
        \includegraphics[width=0.855\columnwidth,height=0.54\columnwidth]{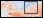}
        \caption{Perturbed prediction}
    \end{subfigure}
    \caption{\textbf{Predicted temperature at the peak of the European Heat Wave 2006.} The forecast is shown $($a$)$ without attack and $($b$)$ after including adversarial observations. The dashed rectangle depicts the target region of the attack.}
    \Description{Two images both showing the outline of western central Europe, in particular the United Kingdom, Germany, France, Denmark and the Benelux states. Both images have a colored overlay showing the predicted temperature on the peak of the European heatwave 2006. The image on the top, that is labeled original prediction shows a region of very high temperatures in the middle across the Benelux states and parts of France and Germany. The image on the bottom, containing a dashed rectangle around most of Germany, shows far lower temperatures, even outside this box.}
    \label{fig:heat-case-study}
\end{figure}

\paragraph{European Heat Wave}
To assess the ability of out attack to conceal extreme temperatures, we apply it to the European Heat Wave 2006, which set temperature records across many Western European countries~\cite{Rebetez2009}. \Cref{fig:heat-case-study} presents the temperature forecasts before and after the introduction of adversarial observations. As in the previous case study, the extreme weather signal is effectively suppressed in the targeted region following the attack.

For this specific attack, we include only the eastern portion of the heat wave as the target region (indicated by the rectangle in \Cref{fig:heat-case-study}). Despite this narrow focus, extreme temperatures are also eliminated from adjacent areas. This highlights another key insight: the impact of adversarial observations extends beyond the targeted geographic region, plausibly removing the entire extreme weather event rather than confining the effect locally.
Furthermore, we observe that the altered forecast significantly overshoots the intended objective of merely hiding the heat wave. Instead, it predicts unnaturally mild temperatures ranging from 5°C to 10°C in the regions surrounding the North Sea. This effect could be mitigated by an attacker by specifying a desired target temperature, instead of minimizing the predicted temperature.

\paragraph{Hurricane Katrina}
Our final case study evaluates the precision with which an adversary manipulate the course of a storm , using Hurricane Katrina as an example.
After initially passing Florida, the storm made its primary landfall near New Orleans~\cite{10.1257/jep.22.4.135}. Rather than suppressing the storm entirely, an adversary may aim to shift the predicted landfall site to disrupt relevant preparations.
To simulate this, we compute adversarial observations that reduce the predicted wind speed at the original landfall site while simultaneously increasing it at a new, perturbed location. 

\begin{figure}
    \centering
    {
        \fontsize{8}{10}\selectfont
        \includegraphics[width=0.855\columnwidth,height=0.4455\columnwidth]{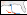}
    }
    \caption{\textbf{Predicted storm path of Hurricane Katrina.} The forecast is shown $($a$)$ without attack and $($b$)$ after including adversarial observations. The triangles show the target location at which the wind speed is minimized (\raisebox{0.12ex}{\tikztriangle[poscolor]{2pt}}) and maximized (\raisebox{0.12ex}{\tikztriangle[negcolor]{2pt}}).}
    \Description{The outline of the United States coast around the Gulf of Mexico with two storm paths depicted by arrows. Both paths start the same location west of the Florida coast and move eastward. The original predicted path coincides with the real storm path, turning north towards the original destination. The perturbed storm path instead continues further eastward, toward the perturbed destination.}
    \label{fig:wind-case-study}
\end{figure}

The original and perturbed storm tracks are shown in \Cref{fig:wind-case-study}. After introducing the perturbation, the forecast storm path clearly deviates from the original, no longer indicating landfall near New Orleans but instead pointing to the manipulated location. The storms trajectory is determined using the location of lowest sea level pressure, which serves as a proxy for the storms eye. Notably, although the optimization process targets wind speed predictions, it also affects atmospheric pressure, again suggesting broader implications of adversarial interference.

%% file: tex/06_countermeasure.tex
Our findings demonstrate that AI-based weather forecasting systems are vulnerable to adversarial observations, underscoring the need for effective defense mechanisms. In the following, we thus explore statistical detection as a potential countermeasure to mitigate this vulnerability, while broader organizational responses are discussed in \Cref{sec:discussion}.

Noisy data is not unique to the adversarial context and in fact a common and practical challenge for real-world forecasting systems. To manage this, quality control procedures are implemented that evaluate the reliability and plausibility of incoming data. These procedures typically consist of hand-crafted rules involving two main categories: whether the observations are temporally and spatially consistent, and whether they fall within a reasonable range of the best estimate of the value~\cite{ens-observations,nwspQC}. Because these checks are designed to handle naturally occurring noise and errors, they are insufficient for detecting the subtle, worst-case perturbations introduced by adversarial observations. We therefore explore whether more sophisticated statistical tests could identify manipulations and serve as a defense against this threat.

We evaluate detecting adversarial observations in the context of a statistical difference to real data. The assimilated state \(\bar{\weatherstate}\) is commonly assumed to consist of an unknown underlying ground-truth value \(\weatherstate\), to which unbiased Gaussian noise is added by the background and observation error~\cite{obsErrors}. We assume a best-case scenario for the defender in which all natural noise can be described by the background error alone. In this setting, the attacker adds noise through the adversarial observations and we arrive at
\[
    \bar{\weatherstate} = \weatherstate + \mathcal{N}(0, \sigma_b^2) + \mathcal{N}(0, \epsilon^2) = \weatherstate + \mathcal{N}\left(0, \sigma_b^2 +\epsilon^2\right),
\]
where $\sigma_b^2$ denotes the variance of the background error.

Under this formulation, any adversarial perturbation increases the total noise in the assimilated state. Thus, if the background error variance is both constant and known exactly, the presence of an attack can, in principle, always be detected---provided the sample size is sufficiently large---since the resulting variance will exhibit a measurable increase.
In practice, however, the sample size is constrained by the number of grid points and the number of variables per grid point, making detection inherently probabilistic. Moreover, the variances of background and observation errors are neither constant nor known with high precision, which makes detecting small increases in noise particularly challenging.

Despite these limitations, we take a conservative approach to evaluate the overall detectability of the attack, assuming a best-case scenario for the defender in which the total error variance in the assimilated state is both constant and known. Under this assumption, we can determine whether a given sample shows a significantly higher variance by applying a simple chi-square test for the variance~\cite{StatisticalMethods}.

\paragraph{Chi-square test setup}
We consider the targets described in \Cref{sec:eval} and compute the minimum increase in noise required to trigger an extreme weather deviation. This is estimated by linearly interpolating the induced deviations across the evaluated noise levels. To ensure that each attack can reach the extreme weather threshold, we do not impose a limit on the maximum noise level. In such cases, we extrapolate beyond the defined maximum attack budget.
For all attacks, we then estimate the detection probability using a chi-square test for variance, assuming perfect knowledge of the expected amount of noise.

\paragraph{Detection results}
The detection probabilities are presented in \Cref{tab:detectability}. Adversarial observations from both baselines are consistently detected using the chi-square test, with rates exceeding $95\,\%$ in all cases---except when temperature is manipulated by the DP-Attacker.
In contrast, our attack results in significantly lower detection probabilities: approximately $\approx 3\%$ for wind speed and temperature, and just $0.2\,\%$ for precipitation.

These results demonstrate that, even under ideal conditions, the attack would likely evade detection. This conclusion is further reinforced by the fact that the assumed detection method is not practically feasible and would likely result in false positives. Consequently, even if such a method were implementable, successful detection would remain unlikely and establishing definitive proof of an attack even more so. We therefore conclude that statistical detection is, unfortunately, not a viable approach for defending against adversarial perturbations in weather forecasting.

\begin{table}[t]
    \caption{\textbf{Detectability of different approaches used to fabricate extreme weather deviations.} The detectability is measured using a chi square test for the variance with best-case assumptions of constant and perfectly known variance of the assimilation error.}
    \label{tab:detectability}
    \centering
    \input{tables/detectability}
\end{table}

%% file: tables/detectability.tex
\begin{tabular}{l rrr}
    \toprule
    \textbf{Method} & \multicolumn{1}{c}{\textbf{Wind Speed}} & \multicolumn{1}{c}{\textbf{Temperature}} & \textbf{Precipitation} \\
    \midrule
    AdvDM  & $>99.99\,$\% & $99.92\,$\% & $>99.99\,$\% \\
    DP-Attacker & $95.04\,$\% & $45.85$\% & $95.33\,$\% \\
    Ours  & $3.07\,$\% & $2.96\,$\% & $0.20\,$\% \\
    \bottomrule
\end{tabular}

%% file: tex/07_discussion.tex
Our findings highlight a critical vulnerability in modern weather forecasting: the integration of machine learning into the prediction pipeline introduces a new attack surface for manipulation. These concerns align with prior research that has revealed fundamental limitations in the robustness of machine learning systems~\cite{madryPGD,kurakinPGD}.
Even more concerning, as demonstrated in \Cref{sec:countermeasures}, such manipulations are likely to remain undetected. While crafting adversarial observations may exceed the capabilities of typical cybercriminals, they represent a promising tool for more sophisticated and well-resourced actors, including nation-state adversaries.
In the following, we thus take a broader perspective on the impact of our work, beginning with a discussion of its limitations and followed by recommendations for mitigating the underlying threat.

\subsection{Limitations}

We begin by outlining the key assumptions underlying our attack and how they may limit its practical impact.

\paragraph{Access to prediction model}
Our attack relies on computing gradients of the model's outputs, which requires access to the model weights. Currently, this is a reasonable assumption, as many leading forecasting models are publicly available~\cite[e.g.,][]{gencast,aardvark,aifs,graphcast}.
However, it is possible that future models will not be publicly released, which would significantly hinder an adversary's ability to carry out the attack. Black-box attacks on machine learning models typically require vastly more queries to the target model~\cite{brendel-18-decision}, rendering such approaches impractical for weather models. This difficulty is further exacerbated by the operational nature of weather forecasting systems, which generally produce predictions only once per time step. For example, conducting 1,000 queries---on the lower end of what is typical for black-box attacks---against a model with a 12-hour time step would require approximately 500 days to complete.

To overcome this constraint, black-box methods would likely need to identify a universal adversarial perturbation that remains effective across multiple time steps. A more feasible alternative arises if the attacker has regular access to the output forecasts of the target system. In this case, a model extraction (or model stealing) attack could be performed, allowing the adversary to reconstruct an approximate surrogate of the target model over time. Adversarial observations could then be crafted using this surrogate in a white-box setting and transferred to the original system. However, model extraction would be slow in this case, as the attacker cannot control the inputs, and thus the process would again require a significant amount of time.

\paragraph{Continuous attack}
In this work, we focus to introduces perturbations at a time step $t$ to manipulate the prediction at a future time step $t+j$.
In practice, however, weather forecasts are updated continuously and new predictions are typically made at each time step. This would require the attack to sustain the manipulations until reaching the forecast at $t+j$. This adds a layer of complexity to the attack, as the adversary must persist with the attack long enough for decisions to be influenced by the forecast.
This persistence does not have to be negative and could  also work in the adversary's favor. Since data assimilation implicitly incorporates the entire history of observations, it may be possible to exploit this process, potentially making the attack more effective. Furthermore, if we extend our view to earlier time steps, smaller adversarial perturbations could be distributed over a longer period, potentially making the attack more subtle and harder to detect. We leave this as an interesting direction for future work.

\paragraph{Problem space}
We consider an attack on the assimilated state on the grid, while an attacker can only control the observations before data assimilation. Although this might seem to constrain our attack to the realm of theoretical feature-space attacks, we have ensured a practical scenario by considering the problem space across all points. The influence of the attacker is realistic in adding only noise and the derived constraint is both conservative and faithful to real-world constraints. Furthermore, our statistical detection is not only inspired by real-world quality control procedures, but assumes a far stronger defender that still cannot reliably detect our attack. Additionally, current developments indicate that data assimilation will also be integrated to achieve end-to-end AI-based weather forecasting in the near future~\cite{DiffDA,aardvark}. This would enable directly computing gradients to the individual observations, allowing attackers to perform the same attack directly on  the problem space.
\subsection{Countermeasures}

Given the limitations of detecting adversarial observations, we consider alternative defense strategies that extend beyond detection.

\paragraph{Selective verification}
A straightforward approach to enhancing forecasting robustness is to cross-verify predictions using traditional numerical weather prediction (NWP) systems whenever extreme weather events are forecast.
This strategy preserves the benefits of shorter runtimes and improved accuracy offered by the MLWP system, while potentially mitigating exposure to adversarial threats. However, this approach alone is not sufficient.

First, such selective verification would fail to detect the second attack scenario, where an adversary suppresses an impending extreme weather event from the forecast since no secondary check would be triggered in the absence of an extreme weather prediction. 
Moreover, even for the first attack scenario, a conflicting forecast from NWP would not necessarily indicate an attack or an error on the part of MLWP, given that MLWP has demonstrated the ability to predict extreme events earlier and with greater accuracy~\cite{gencast}. Consequently, operating MLWP and NWP systems in parallel does not constitute an adequate long-term countermeasure.

\paragraph{Adversarial robustness}
In other domains, adversarial training has proven effective in improving the robustness of machine learning models~\cite{madryPGD,trades}. However, given the complexity and immense computational resources required to train state-of-the-art forecasting models, adversarial training is likely prohibitively expensive or negatively impacts performance relative to traditional prediction systems. We leave a deeper exploration of this approach to future work within the meteorological community.
As a more practical remedy, we recommend that future MLWP development prioritize not only forecast accuracy but also robustness, by systematically evaluating models against attacks, such as ours.
While this strategy may not entirely eliminate the risk of adversarial observations, it could raise the noise threshold required for a successful attack, reducing its impact or making it more likely to be detected through statistical testing.

\paragraph{Trusted data sources}
The existence of adversarial observations highlights a fundamental dependency on the integrity of data sources used in weather forecasting. In safety-critical contexts---such as military or space operations---this dependency necessitates the exclusive use of trusted and rigorously validated observational inputs. Although this constraint may reduce forecast accuracy, it significantly lowers the risk posed by adversarial data. However, such measures cannot entirely eliminate the threat, as a determined adversary may still succeed in compromising individual sources without detection by any trusted entity.

%% file: tex/08_related_work.tex
A substantial body of research has focused on generating adversarial examples for machine learning classifiers~\cite{madryPGD,kurakinPGD,CarWag17,CroHei20}. In contrast, comparatively little attention has been devoted to attacks targeting diffusion models or weather forecasting systems.

\subsection{Attacks on Diffusion Models}
Diffusion models were initially developed and explored in the image domain, where they also faced the first wave of attacks. Initial efforts focused on identifying perturbations that make images unlearnable, aiming to safeguard intellectual property~\cite{ShaCryWenZhe+23}. Subsequent work explored how adversarial examples could be used to prevent imitation or replication of specific artistic styles in generated images. An example is AdvDM by \citet{advdm} that we consider in our evaluation.
The diffusion models targeted by these approaches, however, differ significantly from those used in weather forecasting. In image generation, the models typically denoise a target sample directly, whereas in weather forecasting, they generate sequences of samples autoregressively.

More recently, diffusion models have been extended to domains such as robotic control, which more closely parallels weather forecasting due to its reliance on autoregressive sampling. Attacks in this domain have emerged as well, crafting inputs that disrupt a robot's ability to complete its tasks.
Despite domain-specific variations, the core attack strategies are largely consistent, generally relying on single-step denoising to approximate the inference process. We consider the approach DP-Attacker by \citet{dp-attacker} in our evaluation.
However, our findings reveal that such attacks fail to produce adversarial observations with perturbations small enough to be considered imperceptible or practically effective.

\subsection{Attacks on Weather Forecasting}
Attacks have also been explored in the context of weather forecasting, specially for renewable energy planning~\mbox{\cite{adversarialWind1,adversarialSolar1,adversarialSolar2,adversarialSolar3}}. These studies differ significantly from ours in terms of their threat model. Specifically, they assume that the adversary has direct access to manipulate either the outputs of the forecasting system or the historical data of renewable energy generation.
In contrast, we adopt a more realistic and plausible threat model, where adversarial perturbations are introduced through corrupted observations by a malicious actor. Moreover, prior works are limited in both scope and objective, each focusing on a single forecasting goal within a localized region. Our approach, by comparison, evaluates a broader set of attacker goals spanning global locations.

%% file: tex/09_conclusion.tex
AI-based weather forecasting has attracted increasing attention, with leading meteorological institutions actively exploring the integration of such models into operational forecasting systems. Yet, despite notable advances in model architecture and performance, existing systems lack- safeguards against adversarial manipulation of input data.
In this paper, we demonstrate that  diffusion models---such as those used in GenCast---are susceptible to precisely crafted adversarial observations that can alter extreme weather forecasts without significantly affecting the statistical properties of the input. More broadly, we introduce a novel attack framework for generating adversarial examples targeting autoregressive diffusion models, designed to operate under realistic constraints. 

\section*{Responsible Disclosure}

We have initiated a responsible disclosure process with the GenCast development team. We hope to explore new countermeasures in cooperation with the developers.